\documentstyle[aps]{revtex}
\begin{document}
\draft
\title{Our world as an expanding shell} 
\author{Merab Gogberashvili}
\address{Institute of Physics, Tamarashvili st. 6, 380077 Tbilisi, Georgia \\ 
E-mail: gogber@hotmail.com} 
\date{\today}
\maketitle
\vskip 0.5cm
\begin{abstract}
In the model where the Universe is considered as a thin shell expanding in
5-dimensional hyper-space there is a possibility to have just one scale for a
particle theory corresponding to the Universe thickness. From a realistic model the
relation of this parameter to the Universe size was found.
 \end{abstract}
\vskip 0.3cm
PACS numbers: 04.50.+h, 04.20.Jb
\vskip 0.5cm
\section{Introduction}

In several papers our Universe was considered as a thin membrane in a large 
dimensional hyper-Universe \cite{RS,V,S,BK} (for simplicity we investigate here only
the case of five dimensions). This approach is an alternative to the conventional
Kaluza-Klein picture that extra dimensions are curled up to an unobservable size. It
seems that the model where Universe considered as an expanding bubble in five
dimensions \cite{G} do not contradict to present time experiments \cite{OW} and is
supported by at least two observed facts. First is the isotropic runaway of
galaxies, which for close Universe model is usually explained as an expansion of a
bubble in five dimensions. Second is the existence of a preferred frame in the
Universe where the relict background radiation is isotropic. In the framework of the
close-Universe model without boundaries this can also be explained if the Universe
is 3-dimensional sphere and the mean velocity of the background radiation is zero
with respect to its center in the fifth dimension. 

Usually the confining of matter inside the 4-dimensional manifold is explained as a
result
of the existence of special solution of 5-dimensional Einstein equations. This
trapping has to be gravitationally repulsive in nature and can be produced, for
example, by a large 5-dimensional cosmological constant $\Lambda$, while in four
dimensional space cosmological constant can be hidden \cite{RS,V,S}. 

In previous paper \cite{G} it was shown that in 5-dimensional shell-Universe model
there is a possibility to solve the hierarchy problem, since particles in four
dimensions have effective masses. The only mass scale $m$ corresponds to the other
parameters by the relation             
\begin{equation}
m^3 \sim 1/G \sim \Lambda^{3/2} \sim 1/\epsilon g \sim \epsilon^{-3}~~,
\label{1.1}
\end{equation}  
where G and g are the gravitational constants in five and four dimensions
respectively and $\epsilon$ is the thickness of the universe.

In this paper we want to estimate the only parameter of the theory using observable
cosmological data.

\section{Schwarzschild-like solution}

Consider the Einstein 5-dimensional field equations 
\begin{equation}
R_{AB} - \frac{1}{2}g_{AB}R = \Lambda g_{AB} + 6\pi^2 G T_{AB} 
\label{2.1}
\end{equation}  
and seek for a Schwartzschild-like solution for the shell-Universe in the form: 
\begin{equation}
ds^2 = B(a)dt^2  - A(a)da^2 -a^2d\Omega^2 ~~.
\label{2.2}
\end{equation}  
Here 
\begin{equation}
d\Omega^2 = d\kappa^2  + \sin^2\kappa (d\theta^2 + \sin^2\theta d\phi^2)
\label{2.3}   
\end{equation}
is the 3-dimensional volume element with the ordinary spherical coordinates $\theta$
and $\phi$. The "angle" $\kappa$ corresponds to the radial spherical coordinate by
the formula $r = a \sin\kappa$, while $a$ (scale factor) is the 4-dimensional radial
coordinate.

The components of the 5-dimensional Ricci tensors are:
\begin{eqnarray}
R_{00} = - \frac{B''}{2A} +\frac{B'}{4A}\left(\frac{A'}{A} + \frac{B'}{B}\right) -
\frac{3}{2a}\frac{B'}{A} ~~;\nonumber\\
R_{11} = - 2 +\frac{2}{A} +\frac{a}{2A}\left(\frac{B'}{B} -
\frac{A'}{A}\right)~~;\label{2.4}\\
R_{33} = \sin^2\theta  R_{22} = \sin^2\theta \sin^2\kappa R_{11}~~;\nonumber\\
R_{44} = \frac{B''}{2B} - \frac{B'}{4B}\left(\frac{A'}{A} + \frac{B'}{B}\right) -  
\frac{3}{2a}\frac{A'}{A} ~~,\nonumber
\end{eqnarray}
where primes denote derivatives with respect to $a$.

The Einstein equations (\ref{2.1}) for the outer space ($T_{AB} = 0$) takes the
form:
\begin{eqnarray}
\frac{R_{00}}{2B} + \frac{3R_{11}}{2a^2} + \frac{R_{44}}{2A} = \Lambda
~~;\nonumber \\
- \frac{R_{00}}{2B} + \frac{R_{11}}{2a^2} + \frac{R_{44}}{2A} = \Lambda~~;
\label{2.5}\\
- \frac{R_{00}}{2B} + \frac{3R_{11}}{2a^2} - \frac{R_{44}}{2A} = \Lambda
~~.\nonumber
\end{eqnarray}
Subtraction of the first and third equations gives the solution
\begin{equation}
A = 1/B~~.
\label{2.6}
\end{equation}
Then one can derive from (\ref{2.4})  
\begin{equation}
R_{44} = \frac{(R_{11})'}{2aB}~~.
\label{2.7}
\end{equation}  
Now the only remaining equation to solve in the system (\ref{2.5}) is 
\begin{equation}
R_{11} - \frac{2\Lambda}{3}a^2 = aB' + 2B - 2 - \frac{2\Lambda}{3}a^2 = 0 ~~,
\label{2.8}
\end{equation} 
with the solution
\begin{equation}
B = 1 - \frac{C}{a^2} + \frac{\Lambda}{6}a^2~~.
\label{2.9}
\end{equation}

If we require that at large distances $a$ 4-dimensional Newton's law holds with the
potential
\begin{equation}
\phi = - \frac{G M}{a^2} ~~,
\label{2.10}
\end{equation}
the constant $C$ can be easily expressed in terms of the 5-dimensional gravitation
constant $G$ and the mass of the Universe $M$
\begin{equation}
C = -2G M  
\label{2.11}   
\end{equation}

\section{Equation of motion}

Let us use the thin-wall formalism \cite{BKT,BG} and consider a model of
shell-Universe expanding in a 5-dimensional space-time with outer and inner spaces
\cite{B}
\begin{eqnarray}
ds^2_+ = (1 - 2MG/a^2 + \Lambda_+ a^2)dt^2 - (1 - 2MG/a^2 + \Lambda_+ a^2)^{-1}da^2
- a^2d\Omega^2 ~~,\nonumber\\
ds^2_- = (1 + \Lambda_- a^2)dt^2 - (1 + \Lambda_- a^2)^{-1}da^2 - a^2d\Omega^2 ~~,
\label{3.1}
\end{eqnarray}   
separated by a time-like 4-dimensional thin shell with the metric of the close
isotropic model
\begin{equation}
ds^2 = d\tau^2  - a^2(\tau )d\Omega^2 ~~.
\label{3.2}
\end{equation}
Here $\tau$ is the intrinsic time of the Universe.

Using ordinary formulas of the thin-shell mechanism \cite{BKT,BG} the equation of
motion of the shell-Universe has the form:
\begin{equation}
(\stackrel{.}{a}^2 + 1 - 2MG/a^2 + \Lambda_+ a^2)^{1/2} - (\stackrel{.}{a}^2 + 1 +
\Lambda_- a^2)^{1/2}
= - \sigma a~~,
\label{3.3}   
\end{equation}
where overdots miens derivatives with respect to $\tau$. In (\ref{3.3})
\begin{equation}
\sigma  = \frac{1}{3}lim_{\epsilon \rightarrow 0 }
\int\limits_{-\epsilon}^{+\epsilon}
(6\pi^2GT^0_0 + \Lambda) da 
\label{3.4}  
\end{equation}
corresponds to the intrinsic energy density of the shell-Universe.

The expanding rate of the Universe, which can be found from (\ref{3.3})
\begin{equation}
\dot{a}^2  = - 1 + \frac{\sigma^2}{4}a^2 + \frac{(\Lambda_- - \Lambda_+)^2 -
2\sigma^2 (\Lambda_+ + \Lambda_-)}{4\sigma^2}a^2 + \frac{GM(\sigma^2 + \Lambda_- 
- \Lambda_+)}{\sigma^2a^2} + \frac{G^2M^2}{\sigma^2a^6} ~~, 
\label{3.5}
\end{equation}
depends not only on the energy density of the shell (in 4-dimensional Einstein's
theory
we have only first two terms on right side) but on the properties of outer regions. 

\section{Estimations}

Now from the observed data try to estimate the sickness of the Universe $\epsilon$,
the only parameter of the model. As we noticed above it is possible to hide the
cosmological constant from the 4-dimensional world leaving it in five dimensions
\cite{RS,V,S}. Then  we can consider the Universe as a shell of the dust and the
conservation law for the energy-momentum tensor in 4-dimensions 
\begin{equation}
D_{\nu}T^{\nu\mu} = 0 
\label{4.1}
\end{equation}
gives an equation for $\sigma$
\begin{equation}
\dot{\sigma} + 3\sigma \dot{a}/a = 0 ~~,
\label{4.2}
\end{equation} 
with the solution
\begin{equation}
\sigma = GM/ a^3 ~~.
\label{4.3}
\end{equation} 
It is natural also to think that the only difference in cosmological constants of
inter-outer regions is due to existence of the shell and is proportional to its
energy density
\begin{equation}
\Lambda_{-} - \Lambda_{+} = \sigma g/G ~~.
\label{4.4}
\end{equation}
where $g$ and $G \sim g\epsilon$ are gravitational constants in four and five
dimensions respectively. Recalling the estimation for $\Lambda$ from (\ref{1.1}) and
taking, for example
\begin{equation}
\Lambda = 1/4\epsilon^2 ~~,
\label{4.5}   
\end{equation}
we obtain from the motion equation (\ref{3.5})
\begin{equation}
a\epsilon \sim 1/H^2~~ ,
\label{4.6}   
\end{equation}
where $H = \dot{a}/a \sim 10^{-17} sec^{-1}$ is Hubbl parameter. The present size
$a$ 
of the Universe is unmeasurable, but from the relation (\ref{4.6}) we need 
\begin{equation}
a \geq 10^{60} cm
\label{4.7}   
\end{equation}
in order to have the unobservable thickness of the Universe
\begin{equation}
\epsilon \leq 10^{-2} cm ~~.
\label{4.8}   
\end{equation}

So for the realistic assumptions that the Universe can be considered as an expanding 
shell of the dust in five dimensions and difference of cosmological constants
of outer regions is of order of the Universe energy density we obtain the relation
between the  Universe size and its sickness (\ref{4.6}).


\end{document}